\documentclass[11pt,twoside]{article}
\usepackage{asp2014}

\aspSuppressVolSlug
\resetcounters

\begin{document}

\title{Real-time Data Ingestion at the Keck Observatory Archive (KOA)}

% full name: Bruce Berriman
\author{G. Bruce~Berriman$^1$, M.~ Brodheim$^2$, M.~ Brown$^2$, L.~ Fuhrman$^2$, C.~R.~ Gelino$^3$, M.~Kong$^3$, C.~-~H. Lee$^2$, M.~S.~ Lynn$^3$,  J.~ Mader$^2$, T. ~Oluyide$^3$, M.~A.~ Swain$^3$, T.~ Tucker$^2$, 
A. Laity$^3$, J. Riley $^2$}
\affil{$^1$, Caltech/IPAC-NExScI, Pasadena, CA 91125, USA; \email{gbb@ipac.caltech.edu}}
\affil{$^2$ W. M. Keck Observatory, Kamuela, HI 96743, USA}
\affil{$^3$ Caltech/IPAC-NExScI, Pasadena, CA 91125, USA} 
% remove/add as you need

% remove/add authors as you need

% remove/add as you need

% leave these next few aindex lines commented for the editors to enable them. Use Aindex.py to generate them for yourself.
% first presenting author should be the first entry for bold-facing the author index page-reference
%\aindex{Berriman,~B.}
%\aindex{Author2,~S.}
% remove/add as you need

% leave the ssindex lines commented for the editors to enable them, use Index.py to suggest yourshttps://www.overleaf.com/project/6350d0c625710cde5a8bc376
%\ssindex{FOOBAR!conference!ADASS 2022}
%\ssindex{FOOBAR!organisations!ASP}

% leave the ooindex lines commented for the editors to enable them, use ascl.py to suggest yours
%\ooindex{FOOBAR, ascl:1101.010}
  
\begin{abstract}

Since 2004, KOA has ingested data acquired at the Keck Observatory telescopes in Hawaii at the end of each night's observations: data are prepared at WMKO for ingestion, then transmitted to the archive at NExScI.   For some instruments for which an automated reduction pipeline was available, KOA created reduced products intended for quick-look purposes.  

Since February of this year, KOA began to prepare, transfer, and ingest data as they were acquired in near-real time; in most cases data are available to observers through KOA within one minute of acquisition. Real-time ingestion will be complete for all active instruments by the end of Summer 2022. The observatory is supporting the development of modern Python data reduction pipelines, which when delivered, will automatically create science-ready data sets at the end of each night for ingestion into the archive.  This presentation will describe the infrastructure developed to support real-time data ingestion, itself part of a larger initiative at the Observatory to modernize end-to-end operations. 

During telescope operations, the software at WMKO is executed automatically when a newly acquired file is recognized through monitoring a keyword-based observatory control system; this system is used at Keck to execute virtually all observatory functions. The monitor uses callbacks built into the control system to begin data preparation of files for transmission to the archive on an individual basis: scheduling scripts or file system related triggers are unnecessary. An HTTP-based system called from the Flask micro-framework enables file transfers between WMKO and NExScI and triggers data ingestion at NExScI. The ingestion system at NEXScI is a compact (4 KLOC), highly fault-tolerant, Python-based system. It uses a shared file system to transfer data from WMKO to NExScI. The ingestion code is instrument agnostic, with instrument parameters read from configuration files. It replaces an unwieldy (50 KLOC) C-based system that had been in use since 2004.

Real-time ingestion will enable observers to access their data almost as they are acquired, and positions KOA to support the science of transient, time-varying sources measured with facilities such as Zwicky Transient Factory and the Rubin Observatory (when operational),and to support discovery of electromagnetic counterparts to multi-messenger sources.
  
\end{abstract}

\section{Introduction}

In September 2022, The Keck Observatory Archive (KOA)\footnote{\url{https://koa.ipac.caltech.edu}} began ingesting raw data of all active instruments in near-real time, generally within 1 minute of acquisition. KOA had, from August 2004 to September 2022, ingested data in bulk after completion of the night's observing.

This work is part of a Data Services Initiative (DSI) under way at the observatory to streamline observatory operations. The DSI will tightly integrate observation planning, data acquisition, data reduction, and real-time archiving \citep{2022SPIE12186E..0HB}. In addition to archiving real-time raw data (level 0), KOA will ingest quick-look reduced data (level 1) within 10 minutes of creation at the telescope, and science-grade reduced data (level 2) created automatically at the end of each night. These reduced products are created with modern Python-based Data Reduction Pipelines (DRPs) intended to replace older and sometimes unsupported DRPs, written in a variety of languages. The new DRPs are freely available with Open Source licenses through a GitHub repository \footnote{\url{https://github.com/Keck-DataReductionPipelines}} ; see e.g. \citep{2020JOSS....5.2308P}
Currently, quick-look data are archived in KOA for six instruments, and science-grade data for four instruments. 

Observers have access to their data immediately on ingestion. A dedicated web interface for observers to manage data during the night will be released in February 2023, and a new authentication library will enable observers to share their data with collaborators during the night. All the data become freely available on expiration of the exclusive-use period.

There are substantial benefits to real-time ingestion. Observers and their collaborators, including those providing strategic follow-up for NASA missions, have immediate access to their data through a secure interface. The national and international communities will, for example, be able to exploit Keck data for time-domain and multi-messenger astronomy. Finally, data with no proprietary period will be available to the \emph{entire} astronomical community minutes after they were acquired.

\section{Real-Time Data Ingestion Design: From Observatory to Archive}
The data are prepared for ingestion at the observatory and subsequently transferred to the archive at NExScI through an API running under a Flask micro-framework. What follows summarizes the architecture at each organization.

\subsection{Real-time Ingestion Architecture at WMKO}

 WMKO's Real Time Ingestion architecture detects, processes, and transmits to the archive at NExScI all level 0 and level 1 instrument data in real-time, and level 2 data as they are created on invocation of the DRP the morning after an observation. The code is automatically invoked as soon as the instrument writes a file to disk. The new code is written in Python and includes multi-threading to handle multiple tasks asynchronously. It runs as a monitoring daemon and continually monitors, through the Keck Task Library (which controls, coordinates and monitors all observatory subsystems), instrument-specific  keywords that indicate a new file has been written to disk. Once detected, the monitor spawns an instance of Data Evaluation and Processing (DEP) as a single thread. Each DEP thread is monitored in a queue. Should a thread encounter an error, it is reported in an e-mail; the file can be ingested at a later time.

\subsection{Real-time Ingestion Architecture at NExScI}

 The data are transferred to a dedicated file system mounted at NExScI and shared between WMKO and NExScI; this file system eliminates copying files from a file system at WMKO to one at NEXScI and thus eliminates overhead in data transfer. The ingestion process itself verifies the transfer manifest---that is, that all files expected have been received---and determines whether the ingestion is a re-ingestion of a previous failed ingestion, then validates the transferred metadata, copies the metadata to the database and the data to the file system, and reports the status to WMKO. The process is performed by compact (5 KLOC), component-based Python software, which replaces cumbersome (50 KLOC) C-based code that had been used since 2004. The new code is instrument agnostic, with instrument-specific parameters written in templates, and instrument metadata defined in configuration files. 

\section{Ingestion Performance}

Figures 1 and 2 show histograms of the ingestion times of raw data for eight instruments, and quick-look (level 1) data from KCWI. The histograms show the ingestion times easily meet their requirements of five minutes (with a goal of one minute) for raw data and 10 minutes for quick-look reduced data.

%
% \articlefiguretwo{example.eps}{example.eps}{ex_fig2}{Now there are two of them.  \emph{Left:} % An image from long ago.  \emph{Right:} The same exact thing.}
 \articlefiguretwo{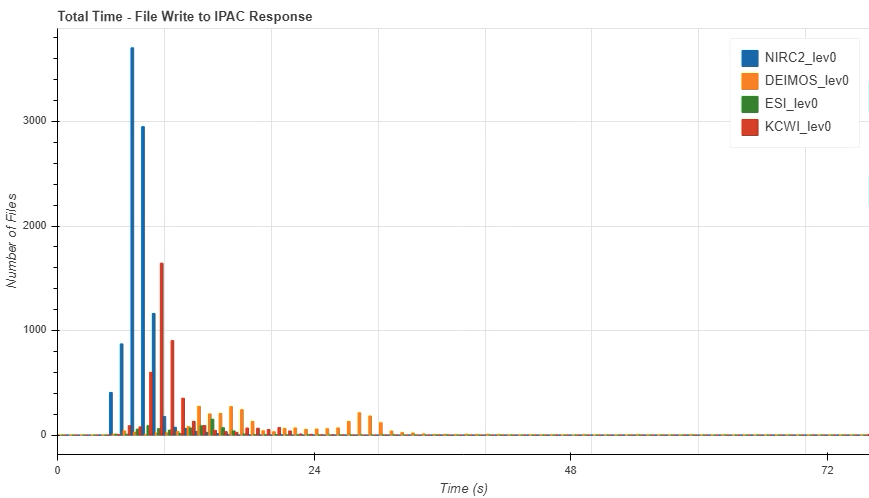}{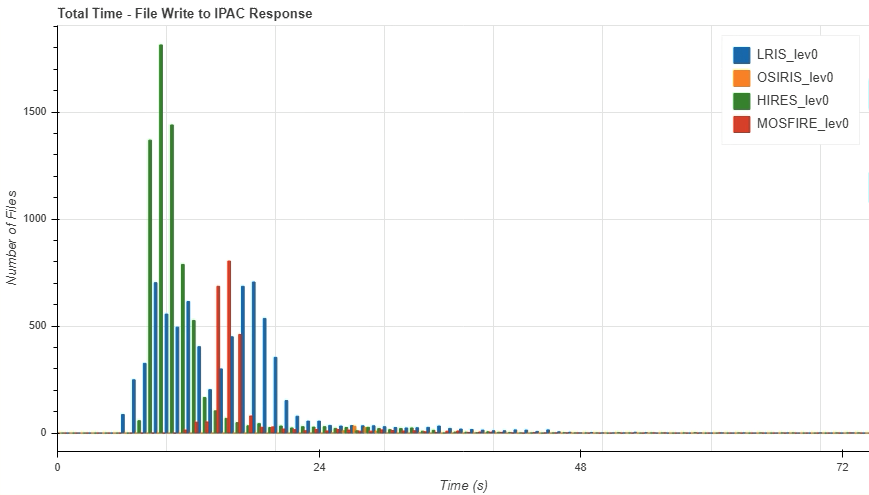}{ex_fig1}{Histograms of the ingestion times of raw data acquired with 8 of the 10 active instruments at the W. M. Keck Observatory. The files are ingested within 1 minute of creation. \emph{Left:} LRIS, OSIRIS, HIRES, MOSFIRE. \emph{right:} NIRC2, DEIMOS, ESI, KCWI }

\articlefigure{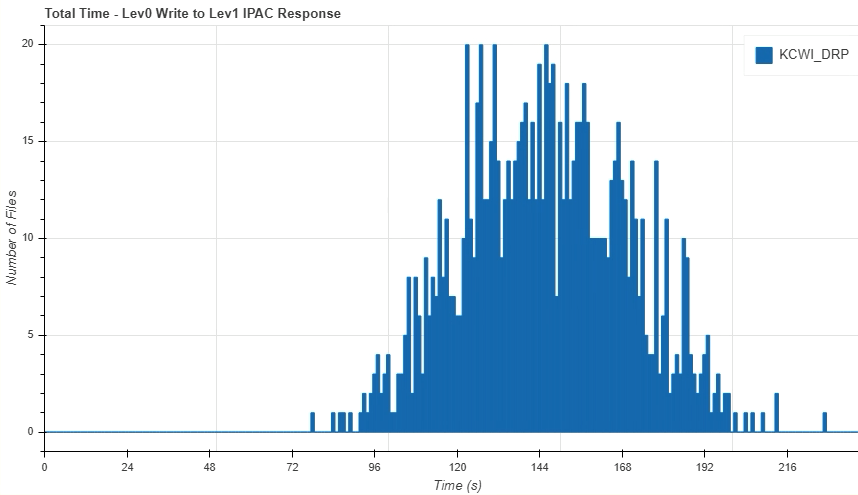}{ex_fig1}{Histogram of the ingestion times of level 1 (quick-look) reduced data for KCWI}

% \articlefigure{Fig3.eps}{ex_fig1}{Welcome to 1953.}
% \clearpage

\section{Future Development}

Over the next year, level 1 and level 2 data will be ingested for the remaining instruments as pipelines are delivered. KOA will, in February 2023, deliver a web-based interface that will enable observers to manage and download their data during the night. It is anticipated that the functionality of the interface will evolve after it enters operations, and that a Python API to support the same functions will be released as well. New instruments commissioned at the observatory will be ingested through the new mechanism described here. The first two such instruments, the Keck Planet Finder (KPF) and the Keck Cosmic Reionization Mapper (KCRM), will be fully commissioned in the first half of 2023. 

% For example in \citet{PID_adassxxx} it was shown that ...

\acknowledgements The Keck Observatory Archive (KOA) is a collaboration between the NASA Exoplanet Science Institute (NExScI) and the W. M. Keck Observatory (WMKO). NExScI is sponsored by NASA's Exoplanet Exploration Program and operated by the California Institute of Technology in coordination with the Jet Propulsion Laboratory (JPL). 

The observatory was made possible by the generous
financial support of the W. M. Keck Foundation.
The authors wish to recognize and acknowledge the very significant cultural role and reverence that the
summit of Mauna Kea has always had within the indigenous Hawaiian community. We are most fortunate to
have the opportunity to conduct observations from this mountain.

\bibliography{P03}

% if we have space left, we might add a conference photograph here. Leave commented for now.
% \bookpartphoto[width=1.0\textwidth]{foobar.eps}{FooBar Photo (Photo: Any Photographer)}

\end{document}